# Peculiarities of Brain's Blood Flow :
# Role of Carbon Dioxide

**Alexander Gersten**

Department of Physics, Ben-Gurion University of the Negev, Beer-Sheva 84105, Israel

**Abstract**

Among the major factors controlling the cerebral blood flow (CBF), the effect of $PaCO_2$ is peculiar in that it violates autoregulatory CBF mechanisms and allows to explore the full range of the CBF. This research resulted in a simple physical model, with a four parameter formula, relating the CBF to $PaCO_2$. The parameters can be extracted in an easy manner, directly from the experimental data. With this model earlier experimental data sets of Rhesus monkeys and rats were well fitted. Human data were also fitted with this model. Exact formulae were found, which can be used to transform the fits of one animal to the fits of another one. The merit of this transformation is that it enable us the use of rats data as monkeys data simply by rescaling the $PaCO_2$ values and the CBF data. This transformation makes possible the use of experimental animal data instead of human ones.

Keywords: Cerebral blood flow, carbon dioxide,

## 1. Introduction

In this work it was found that breathing may have dramatic effects on the brain blood flow. This was already known long time ago to Chinese, Indians and Tibetans. In this paper we will add some simple mathematical models which allow a quantitative description of cerebral blood flow.

In recent years considerable progress was made in utilizing measurements of the regional cerebral blood flow (rCBF) in order to study brain functioning (Knezevic et

al.,1988, Angerson et al.,1989, Costa and Ell, 1991, Howard, 1992). It seems however that the physical and mathematical aspects of the global cerebral blood flow (CBF), or average rCBF, were not sufficiently explored. Our main interest is use of physical principles (Hobbie, 1988), physical and mathematical reasoning as well as means to describe the main features of CBF in a simple way.

The human brain consists of about 2% of the adult body weight, but consumes (at rest) about 15% of the cardiac output (CO) and about 20% of the body's oxygen demand (Sokoloff, 1989, Guyton, 1991).

Glucose is the main source of cellular energy through its oxidation (Sokoloff, 1989, Moser, 1988). The cerebral glucose utilization is almost directly proportional to the CBF , (Harper, 1989, McCulloch, 1988, Harper and McCulloch, 1985). The CBF can be influenced by abnormal glucose levels, is increased during hypoglycemia (Horinaka et al., 1997) and decreased during hyperglycemia (Duckrow, 1995).

Normal mean CBF is approximately 50-55 ml/100g/min, but declines with age (above the age of about 30), in a rate of approximately 58.5-0.24×age ml/100g/min (Maximilian and Brawanski, 1988, Hagstadius and Risberg, 1983, see also Yamamoto et al., 1980, for other details).

The cardiac output can be increased many times (up to about tenfold) during very heavy exercise or work (see Appendix A). Only part of the cardiac output increase can be accommodated by the brain blood vessels because of autoregulatory mechanisms and because of the vessels limited capacitance, which is influenced by their elasticity, limited space of the cranium and the presence of the cerebrospinal fluid (CSF).

Autoregulatory mechanisms exist, which maintain the CBF approximately constant for cerebral perfusion pressure (CPP) over an approximate range of 60-160 mm Hg (Harper, 1989, McCulloch, 1988, Aaslid *et al.,* 1989, Ursino, 1991). Outside this autoregulatory range the CBF may decrease (CPP<60 mm Hg) as in the case of hypotonia (Sokoloff, 1989) or increase (CPP>160 mm Hg) as in the case of high hypertension (Guyton, 1991). Again, the above statements are valid only for normal functioning. For some abnormal functioning the autoregulatory mechanisms may break down, for example if $Pa_{CO_2}$ > 70 mm Hg (Harper, 1989, Harper, 1966).

The CBF is also influenced by the value of cerebral tissue $PaO_2$, whose normal range is about 100 mmHg. Only below approximately 40-50 mmHg there will be a very strong increase of CBF (Guyton, 1991), mobilizing the organism to prevent suffocation.

The main parameter influencing the CBF is the arterial $Pa_{CO_2}$. About 70% increase (or even less) in arterial $Pa_{CO_2}$ may double the blood flow (normal value of $Pa_{CO_2}$ is about 40 mmHg.) (Sokoloff, 1989, Guyton, 1991). A.M. Harper (Harper, 1989) gave an interesting and vivid description of the above situation:

> "However, in the welter of PET scanners, NMRs and SPECTS, with physicists, isotope chemists, computer experts and mathematicians producing reams of data and results accurate to the nth decimal place from pixel sizes which are shrinking by the week, it is easy to forget that a few deep breaths from the patient could lower his arterial carbon dioxide tension by 2 mm Hg (0.27 kPa) and reduce his cerebral blood flow by about 5 per cent. Were this to go unnoticed, the efforts put into our measurements will have been in vain in respect of interpretation of the data for a clinical purpose."

The CBF is very sensitive to $Pa_{CO_2}$ and it is our aim to demonstrate with a simple physical model that important information about CBF capacitance can be obtained by considering only the dependence of CBF on $P_{aCO_2}$. Slowing down the breathing rate, without enhancing the airflow (Fried and Grimaldi, 1993, Fried, 1987, Fried 1990, Timmons and Ley, 1994), or holding the breath, can increase $P_{aCO_2}$. It is plausible that this is one of the essences of yoga pranayama (Bernard, 1960, Iyengar, 1981, Joshi, 1983, Kuvalayananda, 1983, Lysebeth, 1979, Rama, 1986, 1988, Shantikumar, 1987, Shrikrishna, 1996] and of Tibetan six yogas of Naropa (Evans-Wentz, 1958, Mullin, 1996, 1997). It seems that biofeedback training of breathing (Fried and Grimaldi, 1993, Timmons and Ley, 1994), or methods advocated in yoga, may become important for treating health problems.

In Sec. 2 We have developed a simple physical model, and have derived a simple four parameter formula, relating the CBF to PaCO$_2$. With this model, in Sec.3, experimental data sets of rhesus monkeys and rats were well fitted. In Sec.4 exact formulae were found, which allow to transform the fits of one animal to the fits of another one. The merit of this transformation is that it allows to use rats data as monkeys data (and vice versa) simply by rescaling the PaCO$_2$ and the CBF data.

Experimental data on humans available at this time were used for a fit with our model. In Sec. 5. We compare the human data with those of the rhesus monkeys. Sec. 6 includes our conclusions.

## 2. A Mathematical Model of CBF as a function of arterial $CO_2$

Inspection of experimental data, especially the more accurate ones on animals, like those done with rhesus monkeys (Reivich, 1964), or with rats (Sage et al., 1981) led us to conclude that the CBF (which will be denoted later as B) is limited between two values. We interpreted this as follows: the upper limit $B_{max}$ corresponds to maximal dilation of small blood vessels and the lower (non-negative) limit $B_{min}$ to the maximal constriction of the vessels.

Reivich has fitted his data with a logistic model curve (Reivich, 1964), which has two assymptotes

$$B(p) = \left(20.9 + \frac{92.8}{1 + 10570 \exp[-5.251 \; \log_{10}(p/1mmHg)]}\right) ml/100g/\min, \quad (2.0)$$

where p= $PaCO_2$ in mmHg.

We will present a model which will be based on physical assumptions. Instead of the variable $B$ (the CBF) we will use the normalized to unity quantity $z$ defind as

$$z = \frac{B - B_{min}}{B_{max} - B_{min}}. \quad (2.1)$$

The dependence of the CBF on $PaCO_2$ will be described with the dimensionless variable $x=\log(p/p_1)$, where p= $PaCO_2$ in mmHg and $p_1$ is a fixed value of $PaCO_2$, which may be taken to be $p_1$=1mmHg. The variable $p$ is physical only for $p \geq 0$, in order to avoid formulae which may be valid for $p<0$ the variable $x=\log(p/p_1)$, valid for for $p \geq 0$ was introduced.

We can incorporate the above requirements and use the following assumptions:

$$\frac{dB}{dx} = AF(z) = AF\left(\frac{B - B_{min}}{B_{max} - B_{min}}\right), \quad F(1/2) = 1,$$
$$0 \leq z \leq 1, \; 0 \leq B_{min} \leq B \leq B_{max}, \quad (2.2)$$

where A is a constant (reactivity), and $F(z)$ is a function which depends on CBF only. We will add the following boundary conditions on $F(z)$:

$$F(0) = F(1) = F'(0) = F'(1) = 0, \quad (2.3)$$

where $F'(z) = dF(z)/dz$. The condition F(0)=0 corresponds to the requirement that the constriction is maximal at B=$B_{min}$, F(1)=0 correspond to maximal dilation for B=$B_{max}$. Another physical boundary constraint can be formulated for the derivatives $F'(z)$ as follows: $F'(0) = F'(1) = 0$, which means that the approach to the limits is smooth. We will assume that the constricting and dilating forces are the same, mathematically this condition can be expressed in the following manner

$$F(\frac{1}{2}+z) = F(\frac{1}{2}-z), \quad or \quad F(z) = F(1-z). \tag{2.3a}$$

There are many solutions which satisfy the requirements (2.2), (2.3) and (2.3a). We found the following ones:

$$F(z) = 4^n z^n (1-z)^n, \quad and \quad F(z) = \sin^n(\pi z), \quad n \geq 2.$$

We will choose

$$F(z) = \sin^2(\pi z), \tag{2.4}$$

which will also enable us to integrate analytically Eq. (2.2) and to obtain a rather simple result to visualize. We will be able to utilize it for rescaling rats data to rhesus monkeys data and eventually to human data. This path will also enable us to translate rats data to monkey and eventually human data. From Eq.(2.1):

$$dB = (B_{max} - B_{min})dz, \tag{2.5}$$

Eqs. (2.2) and (2.4) can be now converted to

$$\frac{dz}{\sin^2(\pi z)} = \frac{Adx}{(B_{max} - B_{min})}. \tag{2.6}$$

One can easily check that

$$\int \frac{dz}{\sin^2(\pi z)} = -\frac{1}{\pi} ctg(\pi z) + C, \tag{2.7}$$

where C is an arbitrary constant.
From Eq. (2.7)

$$\int_{z_1}^{z_2} \frac{dz'}{\sin^2(\pi z')} = -\frac{1}{\pi} ctg(\pi z_2) + \frac{1}{\pi} ctg(\pi z_1) = \frac{1}{\pi} \frac{\sin \pi(z_2 - z_1)}{\sin \pi z_2 \sin \pi z_1}. \tag{2.8}$$

Integrating Eq. (2.6), using Eq. (2.7), we obtain

$$\int_{z_r}^{z} \frac{dz'}{\sin^2(\pi z')} = -\frac{1}{\pi} ctg(\pi z) + \frac{1}{\pi} ctg(\pi z_r) = \int_{x_r}^{x} \frac{Adx}{\Delta B} = \frac{A(x - x_r)}{\Delta B} \tag{2.9}$$

were $\Delta B = B_{max} - B_{min}$.

From Eq.(2.9) we obtain

$$z = \frac{1}{\pi} \text{arcctg}\left(ctg(\pi z_r) - \pi \frac{A(x-x_r)}{\Delta B}\right) \qquad (2.10)$$

From Eq.(2.10) and the relation

$$arc\ ctg(\pi z) + arc\ tg(\pi z) = \tfrac{\pi}{2}$$

we get:

$$z = \frac{1}{2} - \frac{1}{\pi} \arctan\left(ctg(\pi z_r) - \pi \frac{A(x-x_r)}{\Delta B}\right) \qquad (2.10a)$$

Eq. (2.10a) describes the dependence of CBF (z in Eq. (2.10a)) on $p$ in terms of 5 parameters: $z_r, x_r, A, B_{min}, B_{max}$. The number of parameters can be further reduced to four.

## 2.1 The four parameter formula

A simple way to eliminate $z_r$ from Eq. (2.10a) is to chose $z_r=0.5$ (the value half way between the extremes) then $ctg(\pi z_r)=0$.

Eq. (2.10a) will have the simple form

$$z = \frac{1}{2} + \frac{1}{\pi} \arctan\left(\pi \frac{A(x-x_r)}{\Delta B}\right) \qquad (2.11)$$

and after substituting $z = (B - B_{min})/\Delta B$

$$B = B_0 + \frac{\Delta B}{\pi} \arctan\left(\pi \frac{A(x-x_0)}{\Delta B}\right),$$

and after substituting $x=\log(p/p_1)$, the four parameter formula is

$$B(p) = B_0 + \frac{\Delta B}{\pi} \arctan\left(\pi \frac{A \ln(p/p_0)}{\Delta B}\right), \qquad (2.12)$$

where

$$B_0 = B(z_r = \tfrac{1}{2}) = \tfrac{1}{2}(B_{max} + B_{min}); \quad \left(\frac{dB}{dx}\right)_{B=B_0} = A, \quad x_0 = x(z_r = \tfrac{1}{2}), \quad p_0 = p(z_r = \tfrac{1}{2}).$$

## 3. Experimental data with rhesus monkeys and rats

To our knowledge the experimental data of Reivich (Reivich 1964) are the only published data which tabulate CBF and PaCO$_2$ for individual animals. This will enable us to check our model and find individual variations. The fit results are given in table 1. and displayed in Fig. 1.

**Table 1**

*The fit of Eq. (2.12) to experimental data of individual monkeys (Reivich 1964), all off the monkeys and rats (Sage et al., 1981).*

| Monkey No. | $B_{min}$ ml/100g/min | $B_{max}$ ml/100g/min | $p_0$ mmHg | $A$ ml/100g/min |
|---|---|---|---|---|
| 1 | 44.7±5.2 | 95.2±4.8 | 61.0±2.1 | 147.7±76.6 |
| 2 | | not enough data | | |
| 3 | | not enough data | | |
| 4 | 17.6±2.6 | 125.4±3.3 | 91.4±5.0 | 82.2 ±12.2 |
| 5 | 14.7±2.5 | 123.1±2.8 | 54.6±2.0 | 95.6±11.9 |
| 6 | 9.2±3.1 | 121.1±3.0 | 70.6±4.7 | 48.2±5.6 |
| 7 | 18.3±1.3 | 86.9±2.0 | 43.9±1.7 | 84.7±8.8 |
| 8 | 27.3±2.7 | 142.2±3.7 | 82.7±1.3 | 416±145 |
| all monkeys | 13.6±2.5 | 122.5±2.7 | 60.3±1.9 | 79.3±8.2 |
| rats | 89.0±10.4 | 512.4±18.1 | 38.0±1.1 | 581.7±91.0 |
| all men | 23.3±1.9 | 165.9±3.9 | 53.2±0.7 | 251.5±18.1 |

It should be noted that for monkeys No. 2 and 3 the data were insufficient to determine all the parameters. The error analysis is described in Appendix A, Eqs. (A4), (A14-A16). In Fig. 1 the dotted lines were obtained using the all monkeys parameters. This model seems to describe the main features of the data.

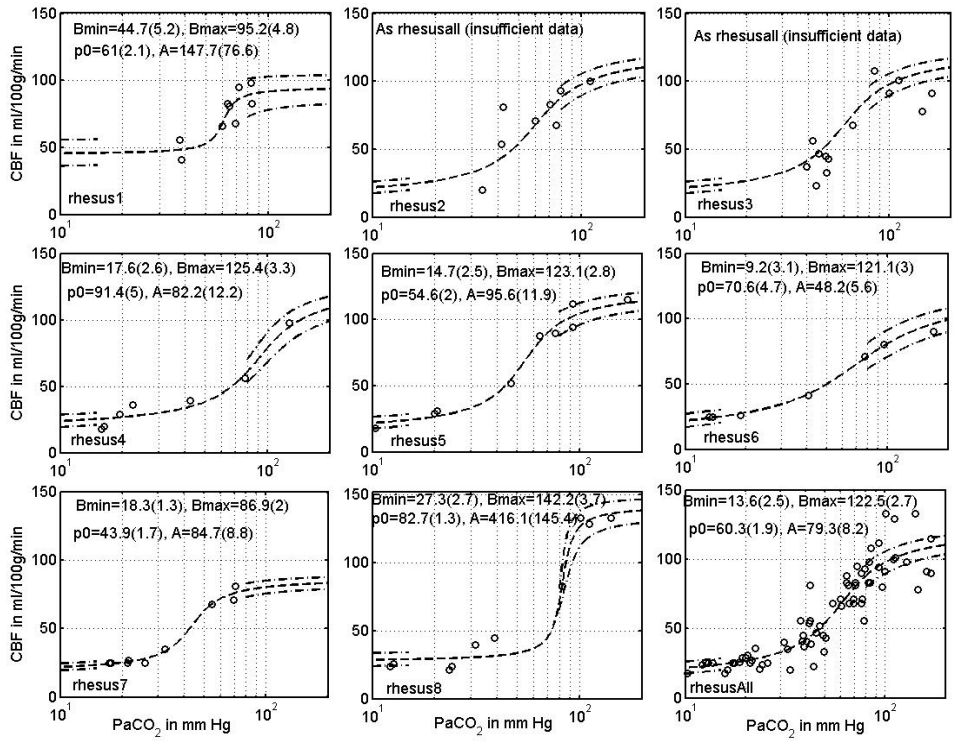

***Fig. 1*** *The fits to the experimental data of (Reivich 1964) with parameters given inTable 1.*
*The dotted line is the rescaling of the monkeys best fit according to Eqs. (4.8) and (4.6).*

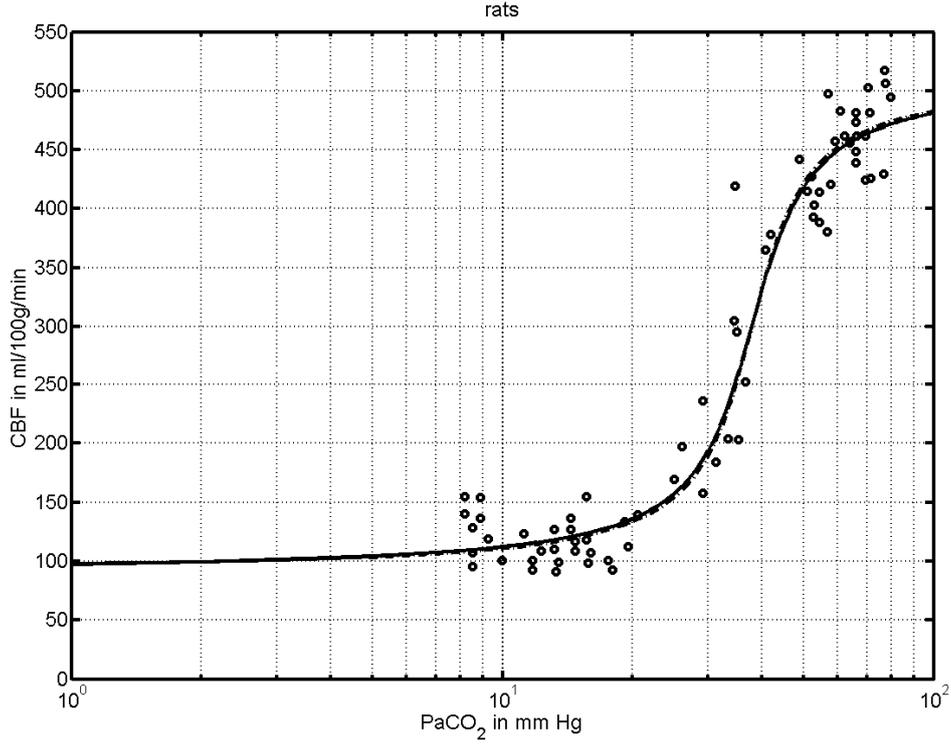

**Fig. 2.** *The dependence of CBF of rats on the partial tension of $CO_2$. The continuous curve is the best fit to Eq. (2.12) for the data of (Sage et al., 1981)]. The parameters are given in Table* 1.

## 4. Rescaling the data

In this section we will transform rat data to serve as monkey data and vice versa. We will utilize Eq. (2.12), and use the upper index R to denote rats and upper index M to denote monkeys. Let us consider two separate fits (e.g. for all monkeys and for rats):

$$B^R(p^R) = B_0^R + \frac{\Delta B^R}{\pi} arctg\left(\pi \frac{A^R \ln(p^R/p^R_0)}{\Delta B^R}\right), \quad (4.1)$$

$$B^M(p^M) = B_0^M + \frac{\Delta B^M}{\pi} arctg\left(\pi \frac{A^M \ln(p^M/p^M_0)}{\Delta B^M}\right) \quad (4.2)$$

Let us assume the possibility that the curve (4.1) is converted to curve (4.2) and vice versa. We will show that it is possible to do this by rescaling the variables p (PaCO2) and B (CBF). Instead of demanding that Eq.(4.1) be equal to Eq. (4.2) we will require the equivalent conditions

$$\pi \frac{A^R \ln(p^R/p^R_0)}{\Delta B^R} = \pi \frac{A^M \ln(p^M/p^M_0)}{\Delta B^M} \quad , \quad (4.3)$$

$$\frac{B^R(p^R) - B^R_0}{B^M(p^M) - B^M_0} = \frac{\Delta B^R}{\Delta B^M}. \qquad (4.4)$$

Equation (4.3) transforms the p coordinates, and Eq.(4.4) transforms the B variables. Thus, for example, if we would like to transfer the monkey data to rat data we will rewrite Eq. (4.3) and Eq. (4.4) as follows

$$p^R = p_0^R \exp\left(\frac{\Delta B^R A^M \ln(p^M / p_0^M)}{\Delta B^M A^R}\right) \qquad (4.5)$$

$$B^R(p^R) = \frac{\Delta B^R}{\Delta B^M}\left(B^M(p^M) - B_0^M\right) + B_0^R. \qquad (4.6)$$

Eq. (4.5) transforms the $p^M$ coordinates (using the parameters of Table 1.) into $p^R$ coordinates and Eq.(4.6) transforms the CBF $B^M$ data into $B^R$ data. The transition of the rat's data to monkey's data is achieved via the equations

$$p^M = p_0^M \exp\left(\frac{\Delta B^M A^R \ln(p^R / p_0^R)}{\Delta B^R A^M}\right) \qquad (4.9)$$

$$B^M(p^M) = \frac{\Delta B^M}{\Delta B^R}\left(B^R(p^R) - B_0^R\right) + B_0^M \qquad (4.10)$$

In Fig. 3. The data of the rhesus monkeys (circles) with the fit of all monkeys (the line as in Fig. 1.) are suplemented with the rat data (stars) according to Eqs. (4.9) and (4.10).

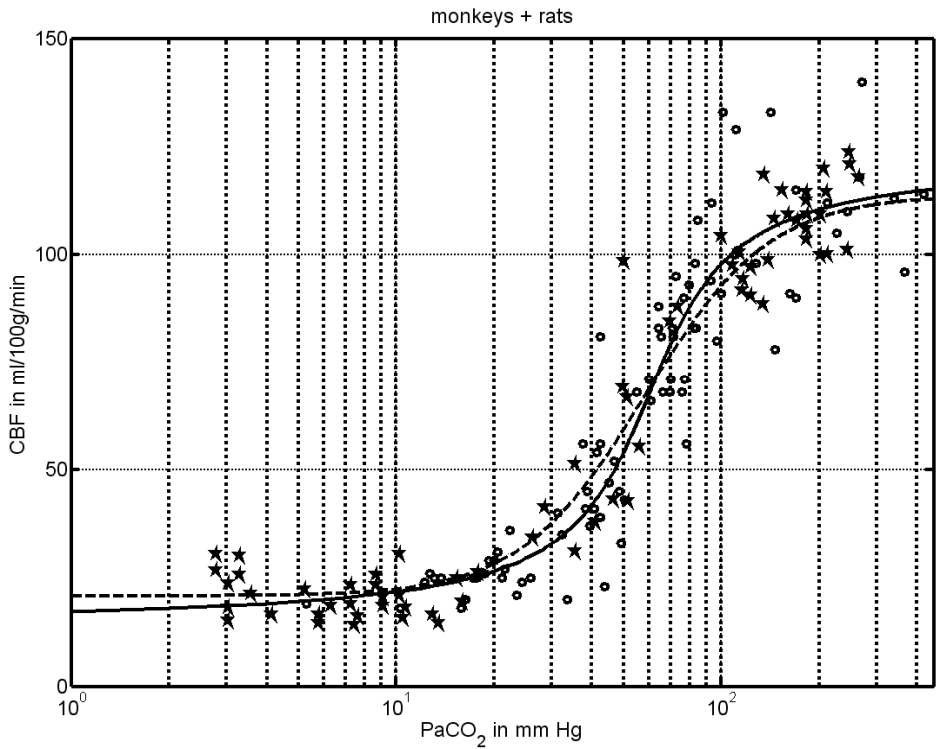

Fig. 3. *The data of the rhesus monkeys (circles) with the fit of all monkeys Eq.(5.2) (solid line) and Eq. (5.1) (dashed line) suplemented with the rat data (stars) according to Eqs. (4.9) and (4.10).*

## 5. Human data

For obvious reasons there are no experimental data on individual humans available in a very wide range of $PaCO_2$. The measurments on animals usually ended with weighing their brains and calibrating the results for 100g of brain tissue. Recently restrictions were imposed on experiments in which animals are killed. In recent years the emphasis was placed on getting regional CBF (rCBF) measurements rather then global CBF. As a result the extended animal measurments are rather old.

Our fit to human data, based on (Reivich 1964, Ketty and Schmidt 1948 and Raichle et all 1970) is given in Fig. 4, with parameters of Table 1 (all men). In (Reivich 1964) it was observed that the human data (in a narrow interval of $PaCO_2$), which existing in that time, were within the experimental errors very close to the rhesus monkey data. One can see it in Fig. 4 (circles, the data of (Reivich 1964)). Therefore the fit of (Reivich 1964), Eq. (2.0)

$$B(p) = \left( 20.9 + \frac{92.8}{1 + 10570 \exp[-5.251 \, \log_{10}(p/1mmHg)]} \right) ml/100g/\min \qquad (5.1)$$

of the rhesus mokey data was and is still being used as the relation between $PaCO_2$ and CBF of humans.

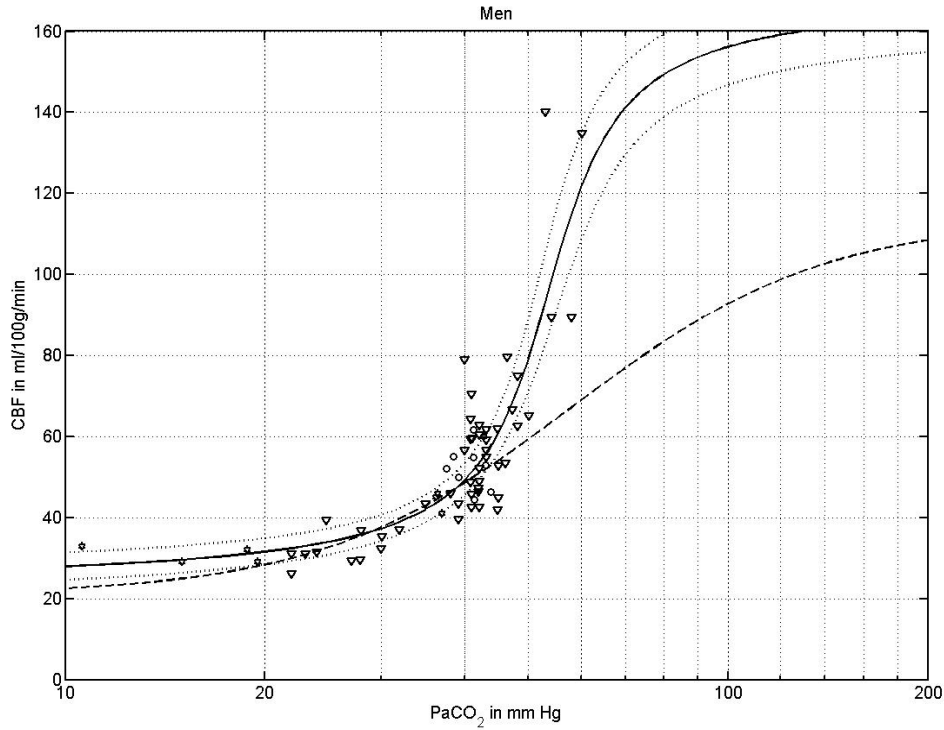

Fig. 4. *Human CBF data (Reivich 1964, circles), (Ketty and Schmidt 1948, triangles) and (Raichle et all 1970, hexagons). The rhesus monkey data are represented by the dashed line of the fit of Eq. (5.1). The dotted lines deviate from the solid line, given by Eq. (5.3), by shifting all parameters by two error bars.*

Our fit to the rhesus monkey data (from Table 1.) is

$$B(p) = 68.0 + 34.7\, arctg\left(2.29\ln(p/60.3 mmHg)\right) ml/100g/\min \qquad (5.2)$$

The fits of Eqs. (5.1) and (5.2) are quite similar and are depicted in Fig. 3. They may serve as a first estimate of human CBF below $PaCO_2$=45 mm Hg. Eq. (5.1) represent the human data in some medical textbooks and publications without mentioning that it is a fit of Rhesus monkeys (for example: Wyngaarden, 1992). Our fit to human data

$$B(p) = 94.6 + 45.4\, arctg\left(5.54\ln(p/53.2 mmHg)\right) ml/100g/\min \qquad (5.3)$$

daviates from that of the Rhesus monkeys in the hypercapnia region ($PaCO_2$ above 45 mm Hg). More accurate human data in this region are needed to confirm our fit.

Changes in $PaCO_2$ induce dramatic changes in CBF. This can be seen in Fig. 5, where relative changes of CBF are given. One can see that changing $PaCO_2$ from normal (40 mm Hg) by 10 mm Hg can change CBF by about 25%-55%.

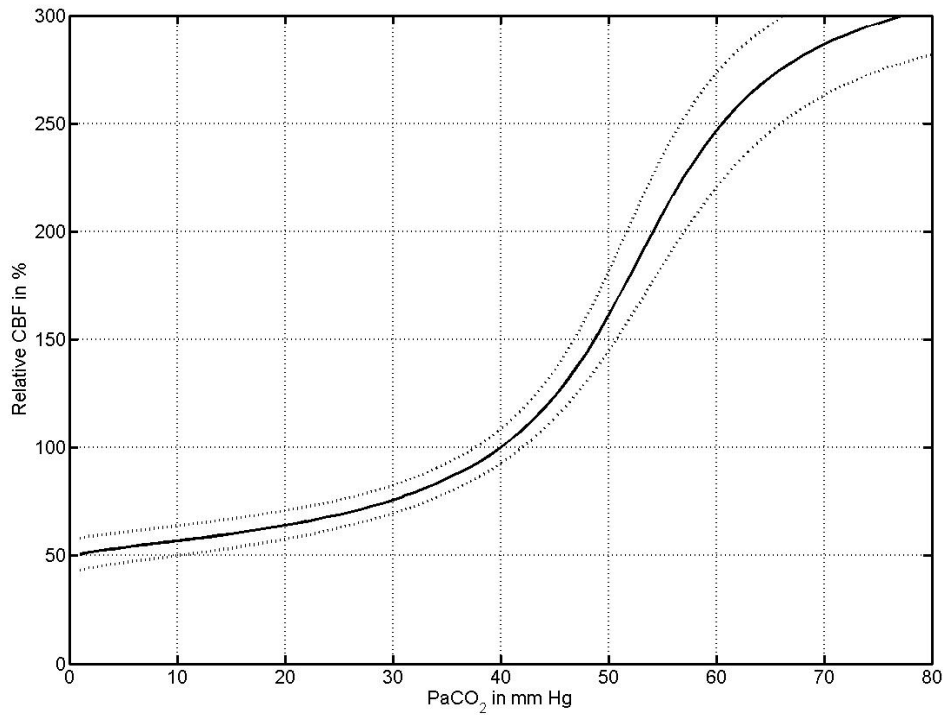

Fig. 5. *Relative changes of CBF with respect to its values at PaCO$_2$=40 mm Hg.*

## 6. Discussion and conclusions

The major factors controlling the cerebral blood flow (CBF) are cerebral perfusion pressure (CPP), arterial partial pressure of oxygen (*PaCO$_2$*), cerebral metabolism, arterial partial pressure of carbon dioxide (*PaCO$_2$*), and cardiac output (CO). The effect of *PaCO$_2$* is peculiar in being independent of autoregulatory CBF mechanisms and allows to explore the full range of the CBF. In Sec. 2 a simple model and a simple formula (Eq. (2.12)), describing the dependence of the CBF on $P_{aCO_2}$ were derived.

The model parameters $B_{max}$, $B_{min}$, A and $p_0$, have a simple meaning and can be determined easily from the experimental data. $B_{max}$ can be associated with the dilation of the blood vessels and with the maximal CBF, therefore it is a good indicator of the extension that the CBF may have. We can expect that it should be age dependent and decline with age. The parameter A, the slope, can be connected with the reactivity of $P_{aCO_2}$. It should also decline with age. Using this formula, we not only found an accurate procedure, but also found age dependent parameters related to the elasticity and adaptability of the blood vessels and CBF.

By analyzing the the age dependence of our model parameters and relating them to the cardiovascular and the respiratory systems one can try to suggest exercises with the aim to improve their values. Experiments are needed to study the age dependence of these parameters.

As the PaCO$_2$ influences strongly the CBF, breathing exercises can be introduced as means to increase the elasticity of brain vessels. For example by alternately hyperventilating and hypoventilating (or breath holding) one can alternately shrink and dilate the cerebral blood vessels, exercising their elasticity.

In theory it is possible to increase *PaCO$_2$* by holding the breath, and in this way increase to large extent the CBF. In practice this is very difficult to achieve, as with a slight change of *PaCO$_2$* from normal, there is a strong urge to ventilate. For very advanced yogis the training of breath holding is of fundamental importance (Bernard, 1960, Rama 1986, 1988). In their opinion, the minimal time to get real benefits is 3 minutes, which is extremely difficult and dangerous for Westerners. It has to be taken into account that the aims of yogis are of a spiritual nature and Westerners can benefit from much milder forms of exercise (Fried and Grimaldi, 1993). Indeed, the milder forms of pranayama (the yoga system of breathing) seem to be quite beneficial, as research around the world indicates (Chandra, 1994, Fisher, 1971, Juan et al., 1984, Kuvalayananda, 1933, Kuvalayananda and Karambelkar, 1957, Nagarathna, 1985, Patel, 1975, Stanescu et a., 1981, Rama et. al., 1979). In Russia the breathing exercises of K.P. Buteyko MD are well known (Buteyko 1983), but they are not documented in advanced scientific journals (outside Russia see Berlowitz, 1995, Hale, 1999). In this method the patients are taught to breathe superficially and to hold (out) the breath for about 1 minute in order to increase their *PaCO$_2$*.

It seems to us that with rhythmic breathing exercises, with some mild breath holding, one can gradually build up the *PaCO$_2$* towards beneficial values, which include improved mental activity.

Although considerale information has already been gained about the effects of *PaCO$_2$* (Kety and Schmidt, 1946, Kety and Schmidt, 1948, Sokoloff et al., 1955, Lassen, 1959, Sokoloff, 1960, Harper and Bell, 1963, Reivich, 1964, 1965, Shapiro et al., 1965, Shapiro et al., 1966, Huber and Handa, 1967, Waltz, 1970, Fujishima et al., 1971, Harper et al., 1972, Paulson et al., 1972, Smith and Wollman, 1972, McKenzie et al., 1979a, 1979b, Yamaguchi et al., 1979, Maximilian et al., 1980, Yamamoto et al., 1980, Sokoloff, 1981), more theoretical and experimental work is needed to clarify this issue.

A special case is that of Dr. Yoshiro Nakamats the greatest modern inventor. In an interview given to Charles Thompson (Thompson 1992) he described the benefits of breath holding to be quite dramatic: " I have a special way of holding my breath and swimming underwater - that's when I come up with my best ideas. I've created a plexiglas writing pad so that I can stay underwater and record these ideas. I call it "creative swimming"... Well, you are shutting off the supply of fresh oxygen and the carbon dioxide content of the blood starts to increase. The body's automatic response is to expand the carotid arteries that feed your brain. They open wide to allow more oxygen rich blood to flow to the brain... By practicing this technique consistently for as little as 3 weeks, you can permanently expand the carotid arteries, so that more oxygen rich blood is flowing to your brain all the time. This activates areas of your brain that suffer from lack of sufficient blood supply and also slows the decay of brain cells. This will translate as a measurable increase in IQ points."

The case of Dr. Yoshiro NakaMats may serve as an example for the possible use of $PaCO_2$ regulation via controlled breathing to enhance brain's blood flow with the aim to improve brain functioning and to prevent brain degenerative diseases. Unfortunately there are not enough experimental data which can shed light on these possibilities. We are initiating experiments in this direction.

In our work we have developed a simple mathematical model which relates CBF to $PaCO_2$. Our model gives good quantitative predictions. Moreover it allows to imitate human data using animal data. With the results of our model it will be possible do devise breating exercises and procedures which aim will be to improve brain's blood circulation. We have already initiated such research, which will be the subject of a separate paper.

# Appendix A
## Error estimation

While performing the best fit (the least-square fit) to the experimental data, the chi squared

$$\chi^2 = \sum_{n=1}^{N} \frac{(t_n - e_n)^2}{(\Delta e_n)^2}, \qquad (A1)$$

is being minimized with respect to the searched parameters. Above, in eq. (A1), $e_n$ are the experimental data, $\Delta e_n$ their errors (or standard deviations), N the total number of experimental points, and $t_n$ are the theoretical predictions for the experimental data $e_n$. The theoretical predictions depend on several parameters which are being searched.

The error of a parameter is defined as the value of the change of this parameter from the best fit, which causes a change in the chi squared by one unit. We will use this procedure throughout the paper, for estimating the errors of the K model parameters $P_k$, k=1,...K. Thus, if near the local minimum at $P_{k0}$ with respect to the parameters $P_k$, the chi squared behaves like

$$\chi^2 = \chi_0^2 + \frac{1}{2}\sum_{k=1}^{K}\sum_{m=1}^{K}(P_m - P_{m0})(P_k - P_{k0})\frac{\partial^2 \chi^2}{\partial P_k \partial P_m}\bigg|_{\chi^2=\chi_0^2} + \text{higher order terms}. \qquad (A2)$$

The error $\Delta P_k$ of the parameter $P_k$ can be evaluated (while all other parameters are at the mimmum) from

$$\chi^2 - \chi_0^2 = 1 = \frac{1}{2}(\Delta P_k)^2 \frac{\partial^2 \chi^2}{\partial^2 P_k}\bigg|_{\chi^2=\chi_0^2}. \qquad (A3)$$

or

$$\Delta P_k = \sqrt{\frac{2}{\frac{\partial^2 \chi^2}{\partial^2 P_k}\bigg|_{\chi^2=\chi_0^2}}} \qquad (A4)$$

The 95% confidence level is achieved with an error of $2\Delta P_k$.

If the experimental errors are unknown special assumptions has to be made. We will consider two casees.

## Case 1 - constant percentage error

Our first assumption is that the errors are proportional to the measured results (i.e., there is a fixed percentage error C):

$$\Delta e_n = C \cdot e_n. \tag{A5}$$

Let $\chi_0^2$ be the minimal value of Eq. (A1). If the theory is exact, the expectation value for the chi squared is

$$\chi_0^2 = N, \tag{A6}$$

otherwise,

$$\chi_0^2 > N. \tag{A7}$$

Substituting eq. (A5) into eq. (A1) and taking into account the two possibilities (A6) or (A7) we obtain:

$$C \leq \sqrt{\frac{1}{N}\sum_{n=1}^{N}\frac{(t_n - e_n)^2}{(e_n)^2}} = C_{UB}, \tag{A8}$$

which gives us an upper bound for the percentage error. Thus the upper bound for the errors of the experimental data (which will be used as an error estimation) will be

$$\Delta e_n = C_{UB} \cdot e_n, \tag{A9}$$

and the estimated chi squared

$$\chi^2 = \frac{1}{C_{UP}}\sum_{n=1}^{N}\frac{(t_n - e_n)^2}{e_n^2}. \tag{A10}$$

The errors of the search parameters will be estimated using Eqs. (A1) up to (A4a).

**Example - linear regression**

As an example, consider the case of a linear fit, with the condition (A5), for a coordinate $x$ and parameters $p_1$ and $p_2$

$$t_n = p_1 x_n + p_2. \tag{A11}$$

The parameters $p_1$ and $p_2$ are obtained by minimizing Eq. (A10) (with an arbitrary constant $C_{UB}$) and afterwards the parameter $C_{UB}$ is obtained using Eq. (A8). The errors of the parameters $p_1$ and $p_2$ will be derived from

$$\chi^2 = \frac{1}{C_{UB}}\sum_{n=1}^{N}\frac{(p_1 x_n + p_2 - e_n)^2}{e_n^2}. \tag{A12}$$

After applying Eqs. (A8), (A10) and (A4) we obtain estimations for the errors

$$\Delta p_1 = C_{UB}/\sqrt{\sum_{n=1}^{N}(x_n/e_n)^2}, \qquad \Delta p_2 = C_{UB}/\sqrt{\sum_{n=1}^{N}(1/e_n)^2} \tag{A13}$$

**Case 2 - constant error**

Here we will assume that the errors are the same for each measurement.

$$\Delta e_n = \Delta e. \tag{A14}$$

Subtituting into Eq. (A1) we obtain

$$\chi^2 = \sum_{n=1}^{N} \frac{(t_n - e_n)^2}{(\Delta e_n)^2} = \frac{1}{(\Delta e)^2} \sum_{n=1}^{N} (t_n - e_n)^2. \tag{A15}$$

If the theory is exact, the expectation value for the chi squared is given by Eq. (A6) and the upper bound for the overall error estimation will be

$$\Delta e = \sqrt{\frac{1}{N} \sum_{n=1}^{N} (t_n - e_n)^2} \tag{A16}$$

The errors of searched parameters can be estimated from Eqs. (A15) and (A4).

**Other constraints**

Other constraints on $\chi^2 - \chi_0^2$ can be worked out. Let us consider the problem of what parameter errors are allowed in order to be within a limited $\chi^2 - \chi_0^2$. Let us outline a general procedure.

Let us introduce the notation

$$M_{km} = \sqrt{\frac{P_{0k} P_{0m}}{2} \frac{\partial^2 \chi^2}{\partial P_k \partial P_m}}\bigg|_{\chi^2 = \chi_0^2} = M_{mk}; \quad u_k = \frac{\sqrt{\chi^2 - \chi_0^2}}{K}, \quad k = 1, 2, \ldots K, \tag{A17}$$

Eq. (A2) can be rewritten as

$$\sum_{k=1}^{K} \sum_{m=1}^{K} \left(\frac{\Delta P_k}{P_{0k}}\right) M_{km} M_{km} \left(\frac{\Delta P_m}{P_{0m}}\right) = \chi^2 - \chi_0^2. \tag{A18}$$

One can note that the solution of the equation

$$\sum_{m=1}^{K} \left(\frac{\Delta P_m}{P_{0m}}\right) M_{km} = u_k \tag{A19}$$

is also a solution of Eq. (A18). One can see it immediately by substituting Eq. (A19) into Eq. (A18)

$$\sum_{k=1}^{K} \sum_{m=1}^{K} u_m u_k = \chi^2 - \chi_0^2.$$

The explicit solutions to Eq. (A19) can be writen in terms of the inverse to the M matrix

$$\Delta P_m = P_{0m} \sum_{k=1}^{K} u_k \left( M^{-1} \right)_{km} = P_{0m} \frac{\chi^2 - \chi_0^2}{K} \sum_{k=1}^{K} \left( M^{-1} \right)_{km} \tag{A20}$$

# **Appendix B**
# **The effect of exercise on cardiac output and CBF**

The highest efficiency for conversion of food energy into muscle work is 25%, the rest is converted into heat (Guyton, 1991). On the other hand the amount of heat produced in the body is directly proportional to the oxygen consumption. At rest the rate of consumption is about 0.2 to 0.3 L/min, and it can increase to 3-6 L/min during maximal exercise Guyton, 1991), depending, among other things, upon age, sex and level of fitness. The energy production is about 5 kcal per 1 L of oxygen consumed. For example, in running, the energy production is approximately 0.2 mL of oxygen per 1 kg of body weight and per meter run. Accordingly, a 70 kg runner, running 2000 m, will produce about 140 kcal of energy. If he will be well insulated with a proper dressing, great part of this energy may be used to increase his body temperature, and for releasing humidity from the lungs. Remembering that 15% of the cardiac output goes to the brain, we may infer that the brain can receive a large amount of heat energy.

According to Hodgkin and Huxley the reaction rate R in the axon changes with the temperature change $\Delta T$ in the following way (Hobbie, 1988): $R = 3^{\Delta T/10^\circ C}$.

The cardiac output C is directly proportional to the work output W and to the oxygen consumption $\dot{V}_{O_2}$. From (Guyton, 1991), which considers typical experimental results, we derived the following linear relations:

$$C = 6.8 + 7(\dot{V}_{O_2} - 0.25 \text{ L/min}) \text{ L/min} , \tag{B.1}$$

$$C = [6.8 + 0.0141 \cdot W] \text{ L/min}, \quad (W \text{ in kg·m/min}) , \tag{B.2}$$

$$\dot{V}_{O_2} = [0.25 + 0.00202 \cdot W] \text{ L/min}, \quad (W \text{ in kg·m/min}) . \tag{B.3}$$

Let us take as an example a runner running 2000m and consuming (additional to the consumption at rest of about .25L/min) 28 L of oxygen. His cardiac output will depend on his speed. For example if he will cover the 2km distance in 14 min. (a mediocre time), we will find from Eq. (A.1) that his cardiac output will be 20.8 L/min, i.e. 3 times larger than his cardiac output at rest. We can see that even a moderate exercise can increase the cardiac output to a large extent. If the body is well insulated, a large amount of energy (in the form of heat) could have been transmitted to the brain. The autoregulatory mechanisms of the brain forbid it from occurring as long as the CPP is below 160 mm Hg. With higher perfusion pressures the CBF and brain's temperature can be increased, but the CBF should be always limited by the value $B_{max}$ which was found in Sec. 2.

According to (McArdle et al., 1996) the CBF increases during exercise by approximately 25 to 30% compared to the flow at rest (Herlhoz et al., 1987, Thomas et al., 1989), but according to (Fox 1999) it can also decrease, especially during heavy exercise. It is easy to understand the reason of these deviations if one takes into account that $PaCO_2$ (the main factor influencing CBF) is proportional to the metabolic production of $CO_2$, but is inversely proportional to the ventilation (Grodins and Yamashiro, 1978, Riggs, 1970). The result depends on the ratio of these two factors, thus CBF may either increase or decrease depending on this ratio.